\documentclass[12pt,preprint,showpacs,preprintnumbers]{revtex4}
\usepackage{amsmath}
\usepackage{graphicx}
\usepackage{dcolumn}
\usepackage{bm}
\usepackage[T1]{fontenc}
\usepackage{ae,aecompl}
\usepackage{epstopdf, epsfig}
\usepackage{color}
\usepackage{appendix}

\makeatletter
\AtBeginDocument{\@ifpackageloaded{natbib}{\ifNAT@numbers\if@filesw\immediate\write\@auxout{\string\global\string\NAT@numberstrue}\fi\fi}{}}
\makeatother
\begin{document}

\title{Collaboration and Competition Between Richtmyer-Meshkov
instability and Rayleigh-Taylor instability}

\author{Feng Chen$^1$\footnote{Corresponding author. E-mail: shanshiwycf@163.com}, Aiguo
Xu$^{2,3}$\footnote{Corresponding author. E-mail:
Xu\_Aiguo@iapcm.ac.cn}, Guangcai Zhang$^2$}
\affiliation{$^1$School of Aeronautics, Shan Dong Jiaotong University, Jinan 250357, China%
\\
$^2$National Key Laboratory of Computational Physics, Institute of
Applied Physics and Computational Mathematics, P.O. Box 8009-26,
Beijing 100088, China \\
$^3$Center for Applied Physics and Technology, MOE Key Center for
High Energy Density Physics Simulations, College of Engineering,
Peking University, Beijing 100871, China}
\date{\today }

\begin{abstract}
The two-dimensional
Richtmyer-Meshkov Instability(RMI) system and the coexisting
system combined with Rayleigh-Taylor Instability(RTI) are
simulated with a multiple-relaxation time discrete Boltzmann
model. It is found that, for the RMI system, the correlation
between globally averaged non-organized energy flux and
nonuniformity of temperature is nearly 1; the correlation between
globally averaged non-organized momentum flux and nonuniformity of
velocity and that between globally averaged thermodynamic
non-equilibrium strength and nonuniformity of density are also
high. In the coexisting system combined with RTI, the
collaboration and competition mechanisms of the two instabilities
are investigated. In the case where RMI dominates, an interesting
interface inversion process is observed. The parameter regions for
RMI dominates and RTI dominates are given. The effects of gravity
acceleration and Mach number on nonequilibrium are carefully
studied, via which the effects of RTI and RMI strengths on the
extent of material mixing are better probed.
\end{abstract}

\pacs{47.11.-j, 51.10.+y, 05.20.Dd \\
\textbf{Keywords:} discrete Boltzmann model/method; nonequilibrium
effect; multiple-relaxation-time; Richtmyer-Meshkov instability;
Rayleigh-Taylor instability} \maketitle

\section{Introduction}

Richtmyer-Meshkov (RM)\cite{rm1,rm2} and Rayleigh-Taylor
(RT)\cite{rt1,rt2} instabilities are key problems in the field of
inertial confinement fusion (ICF), and also exist extensively in
the fields of weapons physics, industrial processes and natural
phenomena\cite{PhysicsReports2017a,PhysicsReports2017b,Xu2016SC-review}.
Therefore, the research is of important practical significance,
and has attracted a wide variety of interests and got rapid
development
\cite{Betti2006,Gupta2010,Sharma2010,Banerjee2011,Orlicz,Mostert,KM2005,ML2007,tbl1,tbl2,chen2011,
LB2011,guo2013,wang2016,IM2016}, to cite but a few.

However, in order for the problem to be tractable, most of these
studies are only a separate study of RM or RT instability. Yet, RM
and RT instabilities might exist simultaneously and interact with
each other. For example, the uneven shape of the supernova remnant
can be shown to result from the combined influence of RM and RT
instabilities\cite{Aschenbach1995}; in the ejecta of astrophysical
planetary nebulae, the internal collisions give rise to RT and RM
instabilities simultaneously\cite{Balick2002}; both RT and RM
instabilities might be involved simultaneously in the deceleration
phase of ICF implosions\cite{Bradley2014}; in the non-premixed
combustion, both the chemically reacting RT and RM instabilities
occur\cite{Attal2015}. So, both RT and RM instabilities should be
taken into account together in order to simulate the practical
applications. But up to now, such studies are still relatively
very little. Matsumoto et al. \cite{Matsumoto2013} found the
synergetic growth of the RT and RM instabilities enhances the
deformation of the jet interface between the relativistic jets and
the surrounding medium. Meshkov and his colleagues
\cite{Meshkov2013} studied the features of turbulent mixing zone
development at an interface accelerated by a non-stationary shock
wave, and in the case of a joint effect of RT and RM instabilities
emerging, the authors found that these instabilities suppress each
other. He et al.\cite{he2016} proposed a new hybrid-drive
nonisobaric ignition scheme of ICF, and discussed the hydrodynamic
instabilities during capsule implosion involving RT instability
and RM instability, and they found that the linear growth rate for
RT instability was significantly reduced, resulting in strong
stabilization effect. Meanwhile, to our knowledge, the above
research are generally based on the Euler, Navier-Stokes (NS)
equations or other macroscopic continuous dynamics models, but
Euler and Navier-Stokes models are not enough to describe the rich
and complex nonequilibrium effects in the RT and RM system.

To overcome this constraint, we resort to the fundamental equation
of non-equilibrium statistical physics, the Boltzmann equation. In
principle, the Boltzmann equation works for flows in the whole
range of Knudsen number (Kn). The Knudsen number is a
dimensionless number defined as the ratio of the molecular mean
free path length to a representative physical length scale. From
this sense, the Boltzmann works for flows with multi-scale
structures, and the Knudsen number can be regarded as a measure
for the continuity of flow. For a non-equilibrium flow, the
Knudsen number can also be regarded as the ratio of relaxation
time ($\tau$) approaching local thermodynamic equilibrium to a
representative time scale ($t^{R}$) in the flow behavior. From
this sense, the Knudsen number can be regarded as a kind of
measure for the extent of Thermodynamic Non-Equilibrium (TNE).
According to the Chapman-Enskog analysis, the Navier-Stokes
equations describe just the corresponding hydrodynamic model of
Boltzmann equation in the continuum limit or when the system is
only slightly deviated from the local thermodynamic equilibrium.
The Navier-Stokes equations are only for the evolutions of
conserved kinetic moments of the distribution function (density,
momentum and energy), while the Boltzmann equation describe the
evolutions of all the conserved and non-conserved kinetic moments.
In fact, the non-conserved kinetic moments supplement the
hydrodynamic quantities in describing the specific non-equilibrium
status of flow.

The Discrete Boltzmann Method (DBM)
\cite{xu2015aps,xu2016,Kinetic2018} developed from the well-known
lattice Boltzmann
method\cite{ss2001,ss1992,xu2012,Yeoman1995PRL,Fang-Qian2004PRE,Qin2005PRE,Shan2016JFM,Shu2015JCP,Shu2007PRE,Zhong2012PRE,Zhong2015,Zhang-Qin2005JSP}
aims to investigate both the Hydrodynamic and Thermodynamic
Non-Equilibrium (HNE and TNE, respectively) behaviors in complex
flows. It can be composed in the levels of Navier-Stokes
equations\cite{yan2013,lin2014,chen2014,gan2015,xu2015pre,lai2016,chen2016,
Lin2016CNF,zhang2016CNF,Lin2017PRE,Lin-2018-CAF,Lin-2017-SR,Xu-2018-FoP},
Burnett equations\cite{Kinetic2018,ZYD-FoP2018,Gan2018,ZYD-2018},
etc., according to the extent of TNE that the model aims to
describe. It has brought some new physical insights into various
complex flows\cite{Kinetic2018,ZYD-2018}. Besides by theory,
results of DBM have been confirmed and supplemented by results of
molecular dynamics\cite{kw2016,kw2017,kw2017b}, direct simulation
Monte Carlo\cite{ZYD-2018,Meng2013JFM} and
experiment\cite{Lin-2017-SR}. In the system containing both
material interface and mechanical interface such as shock waves
and rarefaction wave, non equilibrium characteristics can be used
for the recovery of main feature of real distribution function,
and provide a criterion for discriminating various interfaces,
which can be used in the design of relevant interface tracking
technology\cite{lin2014,chen2014,lai2016,Lin2017PRE}. In a recent
study\cite{chen2016}, the degree of correlation between the
macroscopical nonuniformity and the nonequilibrium strength in the
Rayleigh-Taylor instability are systematically investigated.

Globally speaking,  the research
about non-equilibrium effects in the hydrodynamic instability is
in its infancy.In this paper, the collaboration and competition
relations between RM and RT instabilities are studied
systematically, and the non-equilibrium characteristics of coupled
RT and RM instabilities system are further explored.

The rest of the paper is organized as follows. Section 2 presents
the MRT discrete Boltzmann model. Section 3 presents the
non-equilibrium effects of RM instability system. The
collaboration and competition relations between RM and RT
instability are shown and analyzed in Section 4. Section 5 makes
the conclusion for the present paper.


\section{Discrete Boltzmann model description}

The MRT discrete Boltzmann equation with gravity term reads as
follows
\begin{equation}
\frac{\partial f_{i}}{\partial t}+v_{i\alpha }\frac{\partial f_{i}}{\partial
x_{\alpha }}=-\mathbf{M}_{il}^{-1}\hat{\mathbf{S}}_{lk}(\hat{f}_{k}-\hat{f}%
_{k}^{eq})-g_{\alpha }\frac{(v_{i\alpha }-u_{\alpha
})}{RT}f_{i}^{eq}\text{,}
\end{equation}%
where $f_{i}^{eq}$ is the equilibrium distribution function in the
velocity space, $\hat{f}_{i}$ and $\hat{f}_{i}^{eq}$ are the
particle (equilibrium) distribution functions in kinetic moment
space, $\mathbf{M}$ is the transformation matrix between the
velocity space and the kinetic moment space,
$\hat{\mathbf{S}}=diag(s_{1},s_{2},\cdots ,s_{N})$ is the diagonal
relaxation matrix, $g_{\alpha }$ is the acceleration.

The transformation matrix and the corresponding equilibrium
distribution functions in kinetic moment space (KMS) can be
constructed according
to the seven moment relations. Specifically, the transformation matrix is $\mathbf{M%
}=(m_{1},m_{2},\cdots ,m_{16})^{T}$,
$m_{i}=(1,v_{ix},v_{iy},(v_{i\alpha }^{2}+\eta
_{i}^{2})/2,v_{ix}^{2},v_{ix}v_{iy},v_{iy}^{2},(v_{i\beta
}^{2}+\eta _{i}^{2})v_{ix}/2,(v_{i\beta }^{2}+\eta
_{i}^{2})v_{iy}/2,v_{ix}^{3},v_{ix}^{2}v_{iy},v_{ix}v_{iy}^{2},v_{iy}^{3},(v_{i\chi
}^{2}+\eta _{i}^{2})v_{ix}^{2}/2,(v_{i\chi }^{2}+\eta
_{i}^{2})v_{ix}v_{iy}/2,(v_{i\chi }^{2}+\eta
_{i}^{2})v_{iy}^{2}/2) $. Choosing the Discrete Velocity Model
(DVM) provides us with high flexibility. Here, the following
discrete velocity model is adopted (see Fig. 1):
\begin{align}
\left(v_{ix,}v_{iy}\right) =\left\{
\begin{array}{cc}
\mathbf{cyc}\!:c\left( \pm 1,0\right) , & \text{for }1\leq i\leq 4, \\
c\left( \pm 1,\pm 1\right) , & \text{for }5\leq i\leq 8, \\
\mathbf{cyc}\!:2c\left( \pm 1,0\right) , & \text{for }9\leq i\leq 12, \\
2c\left( \pm 1,\pm 1\right) , & \text{for }13\leq i\leq 16,%
\end{array}%
\right.  \label{dvm3}
\end{align}%
where $\eta_{i}=\eta _{0}$ for $i=1$, \ldots, $4$, and $\eta
_{i}=0$ for $i=5$, \ldots , $16$. The corresponding equilibrium distribution functions in KMS are $\hat{f}%
_{1}^{eq}=\rho \text{,}\; \hat{f}_{2}^{eq}=j_{x}\text{,}\; \hat{f}%
_{3}^{eq}=j_{y}\text{,}\; \hat{f}_{4}^{eq}=e\text{,}\; \hat{f}%
_{5}^{eq}=P+\rho u_{x}^{2}\text{,}\; \hat{f}_{6}^{eq}=\rho u_{x}u_{y}%
\text{,}\; \hat{f}_{7}^{eq}=P+\rho u_{y}^{2}\text{,}\; \hat{f}%
_{8}^{eq}=(e+P)u_{x}\text{,}\;
\hat{f}_{9}^{eq}=(e+P)u_{y}\text{,}\;
\hat{f}_{10}^{eq}=\rho u_{x}(3T+u_{x}^{2})\text{,}\; $ $\hat{f}%
_{11}^{eq}=\rho u_{y}(T+u_{x}^{2})\text{,}\;
\hat{f}_{12}^{eq}=\rho
u_{x}(T+u_{y}^{2})\text{,}\; \hat{f}_{13}^{eq}=\rho u_{y}(3T+u_{y}^{2})%
\text{,}\; \hat{f}_{14}^{eq}=(e+P)T+(e+2P)u_{x}^{2}\text{,}\; \hat{f}%
_{15}^{eq}=(e+2P)u_{x}u_{y}\text{,}$ and $\hat{f}%
_{16}^{eq}=(e+P)T+(e+2P)u_{y}^{2}$, where pressure $P=\rho RT$ and
energy $ e=b\rho RT/2+\rho u_{\alpha }^{2}/2$. At the continuous
limit, the Navier-Stokes equations with a gravity term for both
compressible and incompressible fluids can be obtained.
\begin{figure}[tbp]
\center\includegraphics*[width=0.40\textwidth]{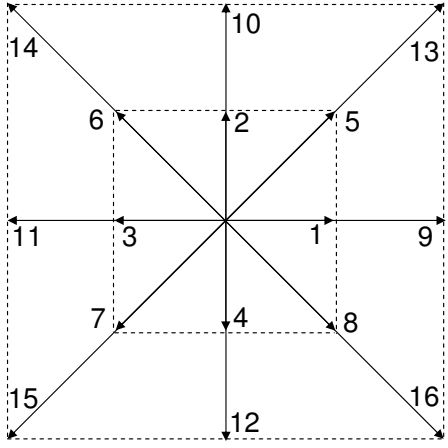}
\caption{Schematic of the discrete-velocity model.}
\end{figure}

The DBM inherits naturally the function of Boltzmann equation,
describing nonequilibrium effects in the system. In the MRT model,
the deviation from equilibrium can be defined as
$\Delta_{i}=\hat{f}_{i}-\hat{f}_{i}^{eq}=\mathbf{M}_{ij}(f_{j}-f_{j}^{eq})$,
and $\Delta _{i}^{\ast }=\mathbf{M}_{ij}^{\ast
}(f_{j}-f_{j}^{eq})$, where $\mathbf{M}_{ij}^{\ast }$ represent
the kinetic central moments, in which the variable $v_{i\alpha}$
is replaced by $v_{i\alpha}-u_{\alpha}$. $\Delta _{i}$ contains
the information of the macroscopic flow velocity, and $\Delta
_{i}^{\ast }$ is only the manifestation of molecular thermal
motion. Corresponding to the simple definition of $\Delta
_{i}^{\ast }$, some clear symbols are more commonly used, that is,
$\Delta_{2\alpha \beta}^{\ast }$, $\Delta _{(3,1)\alpha }^{\ast
}$, $\Delta _{3\alpha \beta \gamma}^{\ast }$ and $\Delta
_{(4,2)\alpha \beta}^{\ast }$, corresponding to $\Delta
_{5,6,7}^{\ast }$, $\Delta _{8,9}^{\ast }$, $\Delta
_{10,11,12,13}^{\ast }$ and $\Delta _{14,15,16}^{\ast }$,
respectively. To provide a rough estimation of Thermodynamic
Non-Equilibrium (TNE), a TNE strength function $d(x,y)$ is
defined,
\begin{equation*}
d(x,y)=\sqrt{\Delta _{2\alpha \beta}^{\ast 2}/T^2+\Delta
_{(3,1)\alpha}^{\ast 2}/T^3+\Delta _{3\alpha \beta \gamma}^{\ast
2}/T^3+\Delta _{(4,2)\alpha}^{\ast 2}/T^4},
\end{equation*}%
where $d=0$ in the thermodynamic equilibrium state and $d>0$ in
the thermodynamic nonequilibrium state. Correspondingly, the
globally averaged TNE strength $D_{TNE}$, Non-Organized Momentum
Flux (NOMF) strength $D_{2}$ and NonOrganized Energy Flux (NOEF)
strength $D_{(3,1)}$ are defined, $D_{TNE}=\overline{d}$, $D_{2}=\overline{\sqrt{%
\Delta _{2\alpha \beta }^{\ast 2}}}$ and $D_{(3,1)}=\overline{\sqrt{%
\Delta _{(3,1)\alpha }^{\ast 2}}}$. A macroscopic nonuniformity
function is also defined as
\begin{equation*}
\delta W(x,y)=\sqrt{\overline{(W-\overline{W })^{2}}},
\end{equation*}
where $W=(\rho, U, T)$ denotes the macroscopic distribution and
$\overline{W}$ is the average value of a small cell around the
point $(x,y)$ \cite{chen2016}.


\section{Non-equilibrium effects of Richtmyer-Meshkov instability}

The DBM model has been validated by some well-known benchmark
tests, and satisfying agreements are obtained between the
simulation results and analytical ones \cite{chen2016}. In this
section, the nonequilibrium characteristics of Richtmyer-Meshkov
instability system are further analyzed. Time evolution is
performed through the third-order Runge-Kutta scheme, and space
discretization is adopted the NND scheme.

An incident shock wave with Mach number $1.2$, traveling from the
top side, hits an interface with sinusoidal perturbation. The
initial macroscopic quantities are as follows:
\begin{equation*}
\left\{
\begin{array}{cc}
(\rho,u_{1},u_{2},p)_{s}=(3.1304,0,-0.28005,2.1187) \text{,} &  \\
(\rho,u_{1},u_{2},p)_{h}=(2.3333,0,0,1.4) \text{,} &  \\
(\rho,u_{1},u_{2},p)_{l}=(1,0,0,1.4) \text{,} &
\end{array}%
\right.
\end{equation*}%
where the subscripts $s$, $h$, $l$ indicate the shock wave region,
the heavy medium region, and the light medium region. The
computational domain is a two-dimensional box with width $W=20.2$
and height $H=80$, and divided into $101\times400$ mesh-cells.
Inflow boundary is applied at the top side, solid is applied at
the bottom, and periodic boundary conditions are applied at the left and right boundaries. $%
\gamma =1.4$ in the whole domain.

Figure 2 shows the evolution of the fluid interface at times
$t=0$, $40$, $60$, and $300$. The parameters are $c=1$, $\eta
_{0}=3 $, $g_{y}=0$, $dt=10^{-3}$, $s=10^{3}$. When the shock wave
passes the interface, a reflected rarefaction wave to the top and
a transmission wave to the bottom are generated, and the
perturbation amplitude decreases with the interface motion. Then,
the peak and valley of initial interface invert, the heavy and
light fluids gradually penetrate into each other as time goes on,
the light fluid rises to form a bubble and the heavy fluid falls
to generate a spike. The transmission wave reaches the solid wall
and reflects, encounters the interface again, converts into a
transmission wave penetrating into the heavy fluid zone and a
reflection wave back to the light fluid. The disturbance of the
interface continues to grow, eventually forming a mushroom shape.
\begin{figure}[tbp]
\center\includegraphics*%
[bbllx=0pt,bblly=150pt,bburx=600pt,bbury=630pt,width=0.6\textwidth]{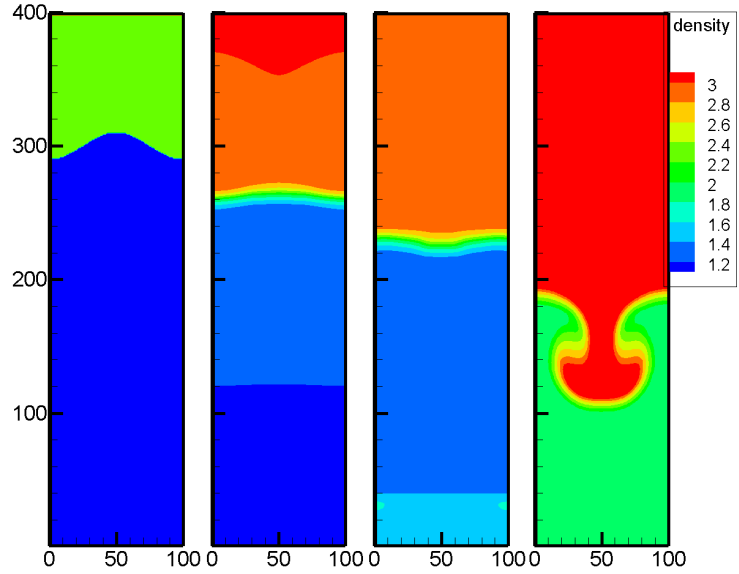}
\caption{Evolution of the fluid interface from a single-mode
perturbation.}
\end{figure}

Figure 3 shows snapshots of the TNE strength $d$ at the same
times. In the position far from the perturbation interface, $d$ is
basically $0$. Around the interface, $d$ is greater than zero. In
the figure, we can find clearly the position of the material
interface and the shock interface. The peak values of TNE strength
$d$ at the shock wave interface are larger than the material
interface (and the rarefaction wave, obviously). This is because
the shock dynamic procedure is faster than the other two
processes, and the system has less time to relax to its
thermodynamic equilibrium. This is consistent with the results in
\cite{lin2014,chen2014}.
\begin{figure}[tbp]
\center\includegraphics*
[bbllx=0pt,bblly=150pt,bburx=600pt,bbury=630pt,width=0.6\textwidth]{%
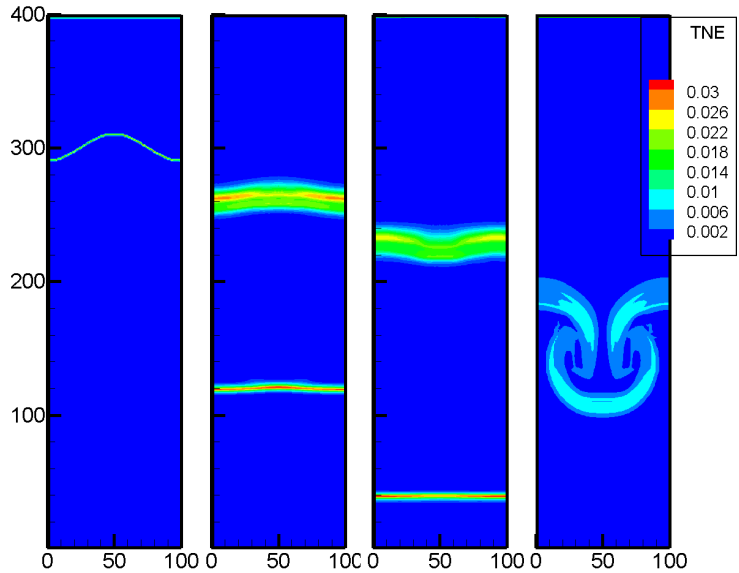} \caption{Snapshots of TNE strength $d$ at times $t=0$,
$40$, $60$ and $300$.}
\end{figure}

Figure 4 shows the degrees of correlation between macroscopic
nonuniformities and various globally averaged nonequilibrium
strength values in the RM system. $\delta \rho$, $\delta T$, and
$\delta U$ are density nonuniformity, temperature nonuniformity,
and velocity nonuniformity, respectively. The degree of
correlation between the temperature nonuniformity and the globally
averaged NOEF strength $D_{(3,1)}$ is approximately $1$.
The correlation between globally averaged NOMF
strength and nonuniformity of velocity and that between globally
averaged TNE strength and nonuniformity of density are also high.

\begin{figure}[tbp]
\center\includegraphics*
[bbllx=0pt,bblly=10pt,bburx=585pt,bbury=310pt,width=0.8\textwidth]{%
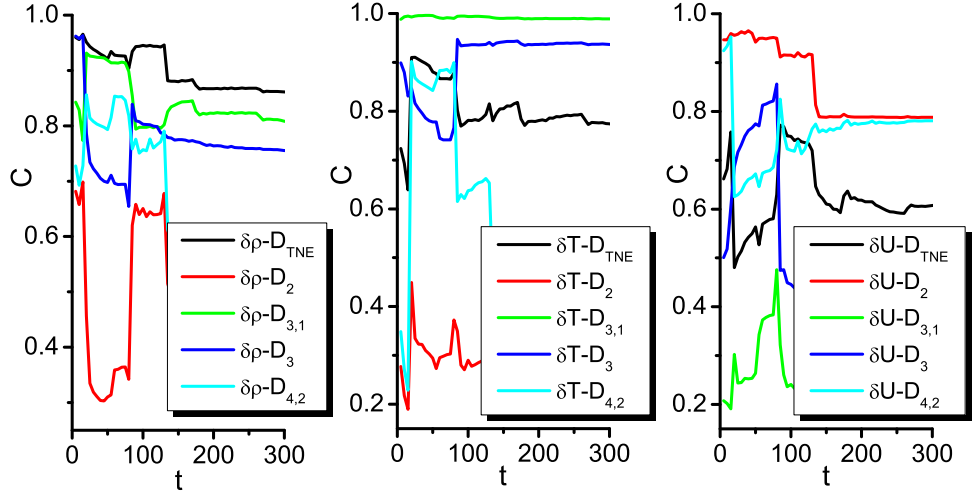} \caption{Degrees of correlation between the macroscopic
nonuniformities and various globally averaged nonequilibrium
strength values. }
\end{figure}

In Figure 5, we can see that, the heat conduction plays a major
role in the degrees of correlation. The greater the heat
conduction, the higher the degree of correlation. Black beelines
are the linear approximation of corresponding curves, that is,
$C=C_{0}+k\times t$. Choosing the proper slope $k$, the value of
$C_{0}$ will increase exponentially with the
increase of heat conduction, and gradually tends to 1.
\begin{figure}[tbp]
\center\includegraphics*
[bbllx=0pt,bblly=10pt,bburx=520pt,bbury=375pt,width=0.8\textwidth]{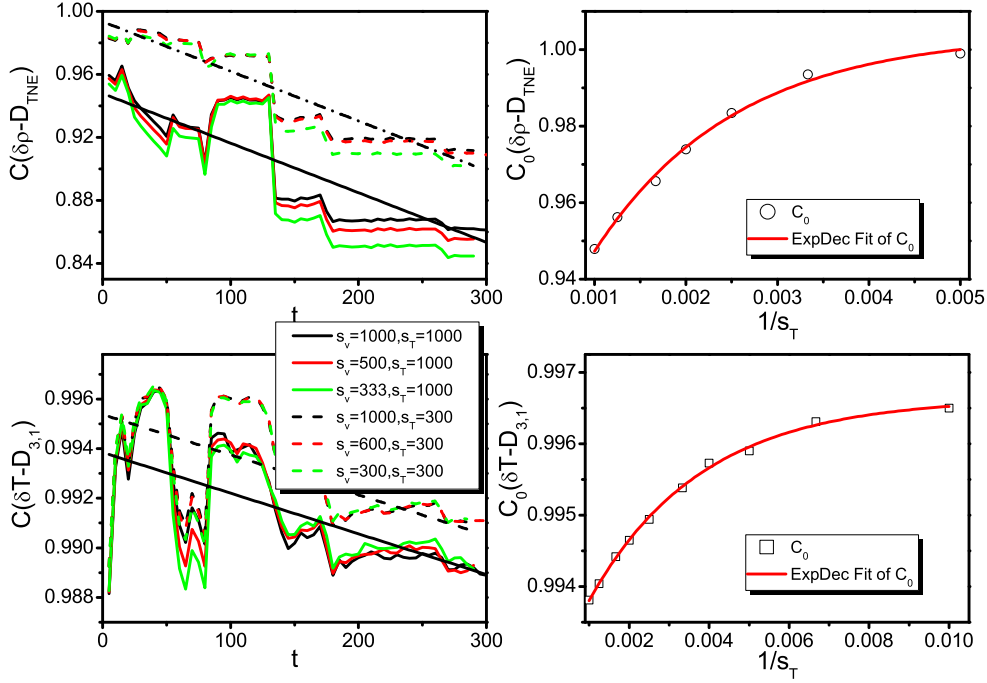}
\caption{Effect of heat conduction on the degree of correlation.}
\end{figure}

\section{Collaboration and Competition Between RM instability and RT instability}

In the section, the RM instability in gravitational field is
simulated, and the collaboration and competition relations between
RM and RT instability are studied. The initial condition is
replaced by the following macroscopic quantities:
\begin{equation*}
\left\{
\begin{array}{cc}
(\rho,u_{1},u_{2},T)_{s}=(3.1304\exp(-g(y_{0}-y_{s})/0.6,0,-0.28005,0.676796) \text{,} & y_{0} < y \\
(\rho,u_{1},u_{2},T)_{h}=(2.3333\exp (-g(y-y_{s})/0.6,0,0,0.6) \text{,} &  y_{s} < y < y_{0}, \\
(\rho,u_{1},u_{2},T)_{l}=(\exp (-g(y-y_{s})/1.4,0,0,1.4) \text{,}
& y < y_{s},
\end{array}%
\right.
\end{equation*}%
where the Mach number of incident shock wave is  $1.2$, $y_{s}$ is
the initial small perturbation at the interface.

\begin{figure}[tbp]
\center\includegraphics*
[bbllx=50pt,bblly=20pt,bburx=510pt,bbury=360pt,width=0.5\textwidth]{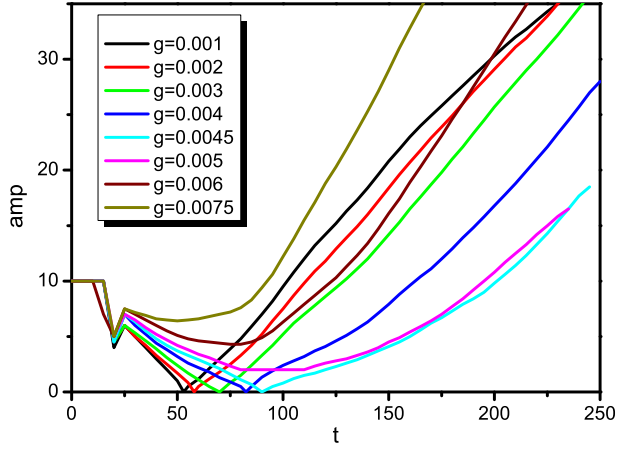}
\caption{Interface amplitude $A$ in different gravity fields.}
\end{figure}
\begin{figure}[tbp]
\center\includegraphics*
[bbllx=68pt,bblly=170pt,bburx=600pt,bbury=480pt,width=0.65\textwidth]{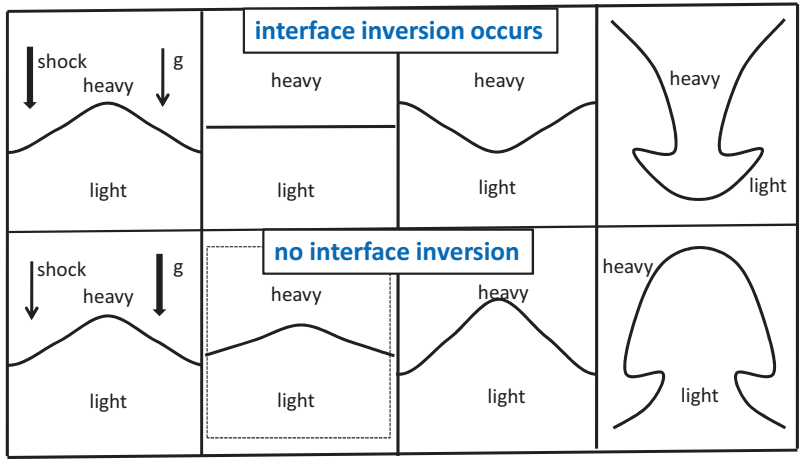}
\caption{The schematic diagram of the collaboration and
competition relations between RM instability and RT instability.}
\end{figure}

Figure 6 shows the interface amplitudes of different gravity
fields. The interface inversion process is affected observably by
the gravity field. There is a competition between the RM
instability and RT instability, and the schematic diagram of the
collaboration and competition relations are shown in Fig.7. The
first line is corresponding to the case of weaker gravitational
action. In the case, the RM instability plays a major role, and
the interface inversion still occurs, but the inversion process is
delayed. After the reversal of the interface, the gravity and
shock wave jointly promote the growth of spike. The second line is
the case of stronger gravitational action relative to shock wave.
In the case, the RT instability plays a major role, and the
interface inversion is suppressed. The light fluid rises to form a
bubble and the heavy fluid falls to generate a spike. The dashed
part indicates the interfacial compression process under shock
wave, and the process does not occur when the impact effect is
zero. The small corner at time $t=20$ in figure 6 indicates the
process. The interface evolution of $g=0.002$ and $g=0.006$ are
shown in Figure 8 and 9, respectively. We specifically explore the
conditions for the interface reversal. The relationship between
the Mach number and the gravitational acceleration is obtained, as
shown in Figure 10. The abscissa is the Mach number, and the
ordinate is the dimensionless gravitational acceleration. The
trend can be expressed by a simple linear function
$g_{nond}=-0.64+0.59Ma$, where $g_{nond}$ is a dimensionless
acceleration of gravity. The region below the fitting curve is the
interface inversion region, i.e., RM instability plays a major
role. In the region above the fitting curve, the interface
inversion mechanism is suppressed and RT instability plays a major
role. The growth of disturbance interface near the fitting curve
is suppressed farthest.

\begin{figure}[tbp]
\center\includegraphics*
[bbllx=5pt,bblly=200pt,bburx=620pt,bbury=580pt,width=0.6\textwidth]{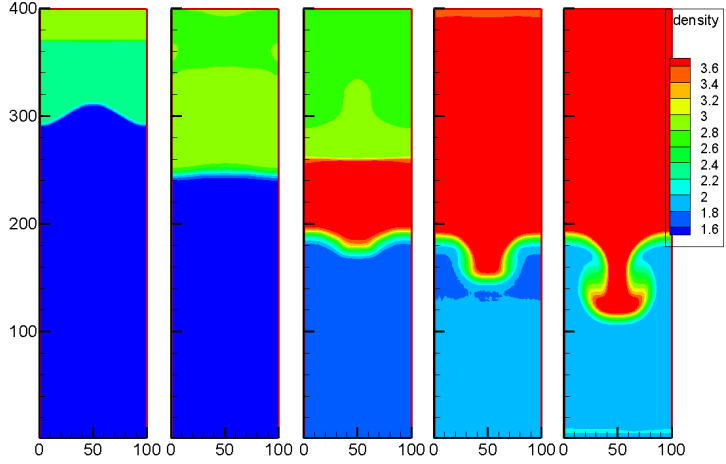}
\caption{The interface evolution of $g=0.002$ and $Ma=1.2$.}
\end{figure}
\begin{figure}[tbp]
\center\includegraphics*
[bbllx=5pt,bblly=200pt,bburx=620pt,bbury=580pt,width=0.6\textwidth]{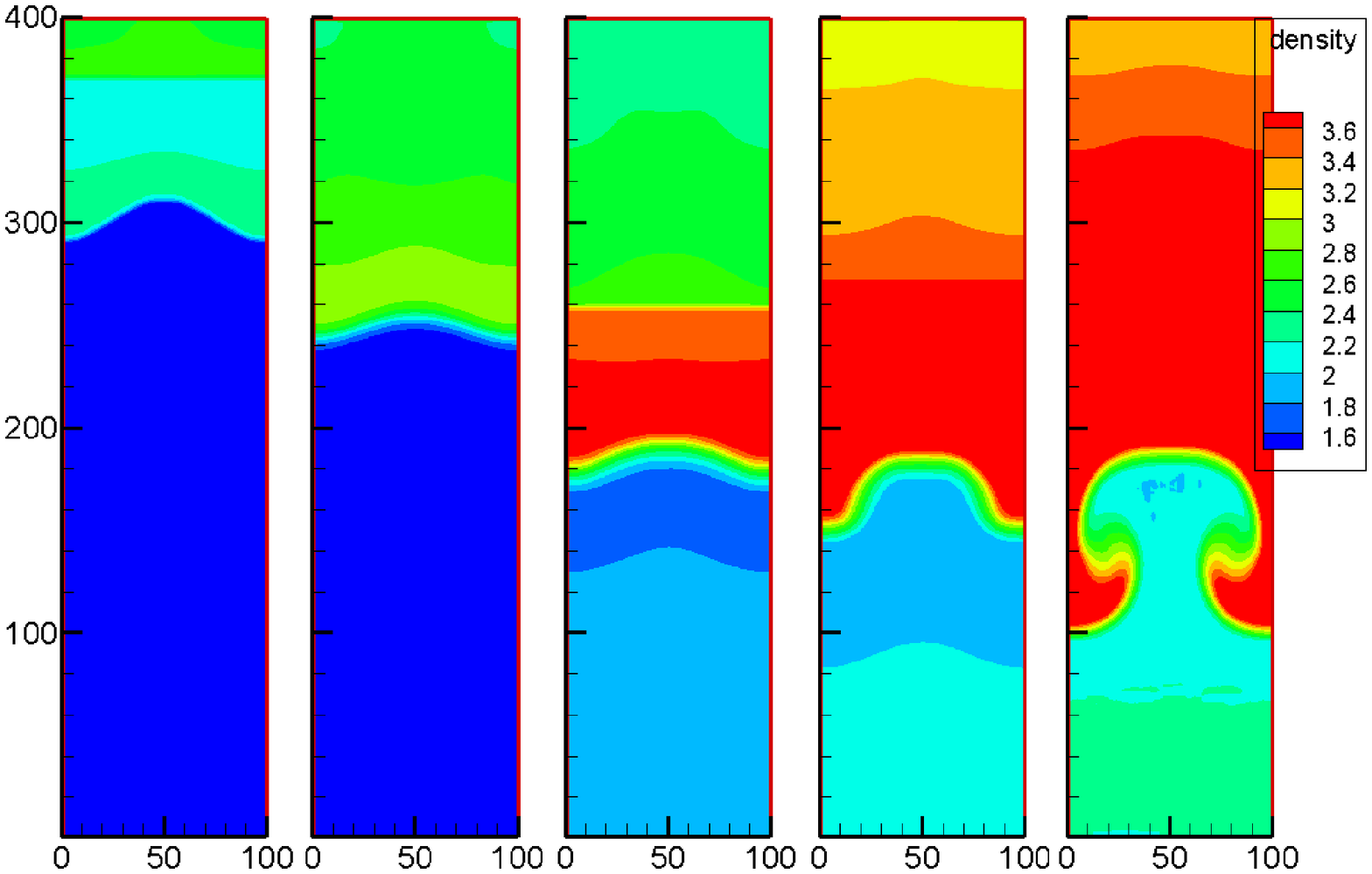}
\caption{The interface evolution of $g=0.006$ and $Ma=1.2$.}
\end{figure}
\begin{figure}[tbp]
\center\includegraphics*
[bbllx=40pt,bblly=10pt,bburx=500pt,bbury=360pt,width=0.5\textwidth]{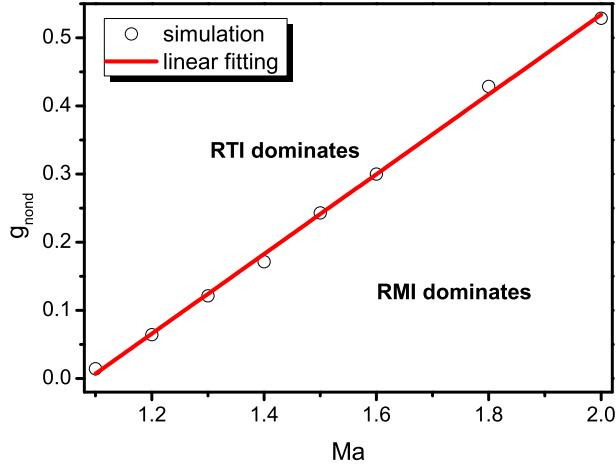}
\caption{The regions for interface inversion or not.}
\end{figure}

Figure 11 shows the thermodynamic nonequilibrium characteristics
of the RT and RM instability coexisting system with different
gravity fields. The jump at time $t=85$ is due to the reshock
process. That is, the transmission shock wave reflects from the
solid wall on the bottom and encounters the interface again.
Before the reshock process, the greater the gravity $g$, the
greater the globally averaged TNE strength $D_{TNE}$. This is
because the larger the gravity $g$, the greater the density
gradient. Both $D_{TNE}$ and $D_{2}$ have large attenuation at
time $t=140$, which is due to the reflected shock wave reaching
the upper boundary. The development of $D_{(3,1)}$ is mainly
affected by the interface. During the process, the interface
continues to grow, so $D_{(3,1)}$ has no attenuation.
Subsequently, for the cases where interface inversion still occurs
($g<0.005$), before the development of Kelvin-Helmholtz (KH)
instability ($t=170$), the smaller the gravity, the larger the
values of $D_{TNE}$ and $D_{(3,1)}$. This is because the greater
the gravity, the greater the resistance to inversion, making the
interface smaller at this time. After the development of KH
instability, the greater the gravity, the greater the growth rate
of $D_{TNE}$ and $D_{(3,1)}$.
\begin{figure}[tbp]
\center\includegraphics*
[bbllx=10pt,bblly=25pt,bburx=545pt,bbury=415pt,width=0.8\textwidth]{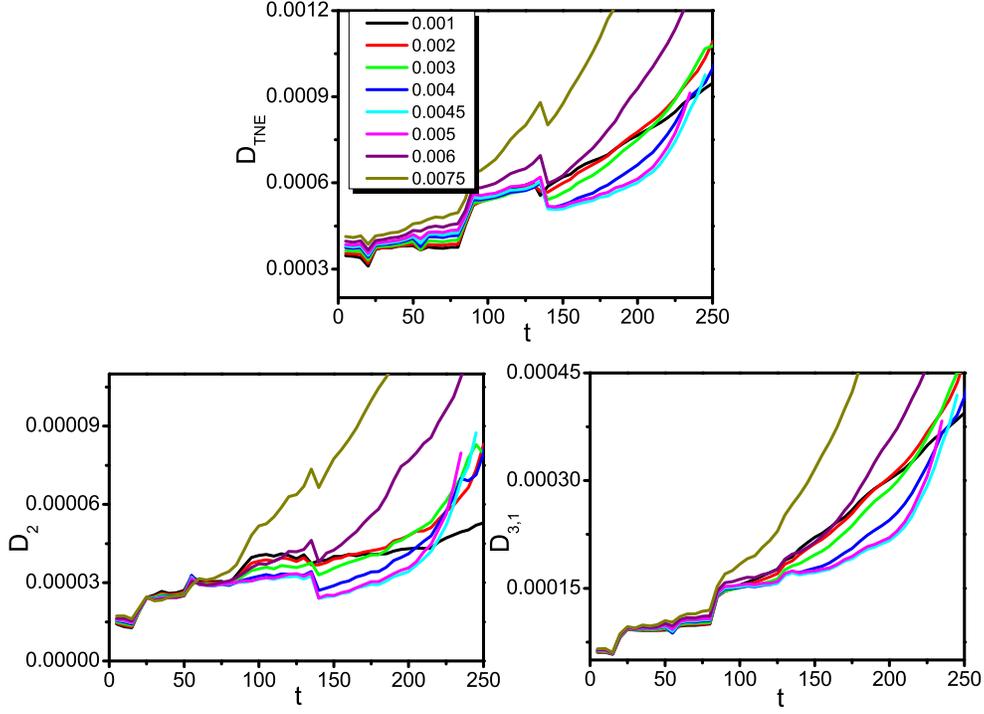}
\caption{The thermodynamic nonequilibrium characteristics of the
RT and RM instability coexisting system with different gravity
fields.}
\end{figure}

In figure 12, we study the effect of Mach number on the
non-equilibrium degree. Under different Mach numbers, the
evolution speeds of the system are different. In order to
guarantee the reliability of the analysis, the systems should have
a similar degree of evolution. For convenience, we adopt the
unperturbed interface, and choose the system that the transmission
shock wave just reaches the solid wall as the research object.
Figure (a) shows the globally averaged TNE strength with different
Mach number. With the increase of Mach number, the non equilibrium
degree of the system is increased. In figure (b), circles indicate
the values of $D_{TNE}$ when the shock wave with different Mach
number reaches the bottom. Numerical fitting shows that the
nonequilibrium degree increases exponentially with the increase of
Mach number. (c) and (e) show the degrees of correlation between
$\delta \rho$ and $D_{TNE}$, and between $\delta T$ and
$D_{(3,1)}$, respectively. The dashed lines are the Origin fitting
results of the corresponding real curves, and the corresponding
values are represented by triangulars in (d) and squares in (f).
It is shown that the degree of correlation almost exponentially
decreases with the increase of Mach number.
\begin{figure}[tbp]
\center\includegraphics*
[bbllx=0pt,bblly=20pt,bburx=520pt,bbury=380pt,width=0.8\textwidth]{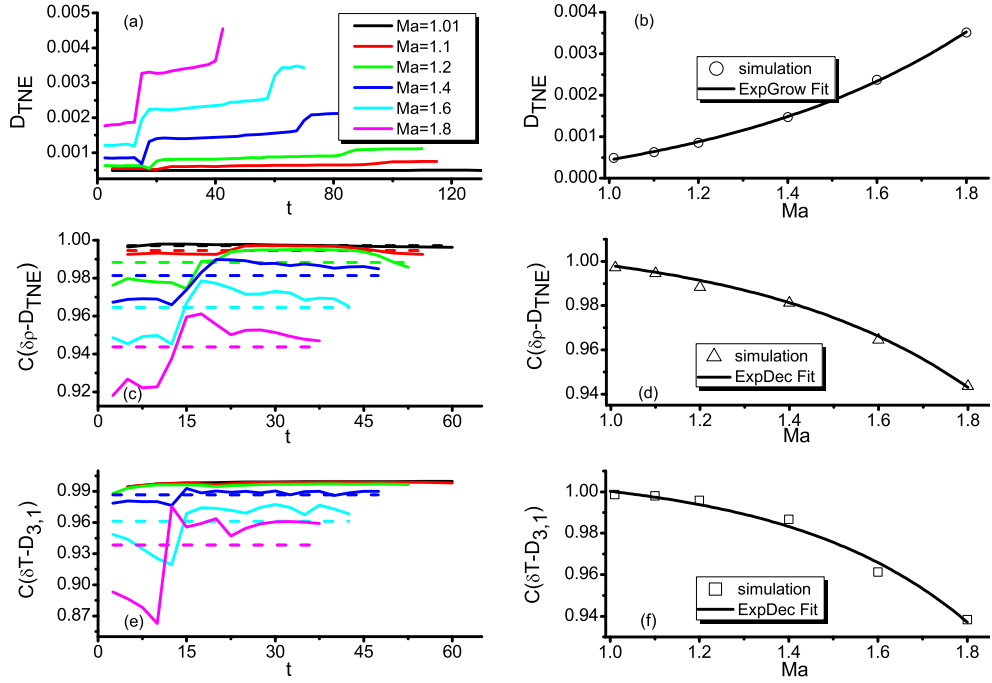}
\caption{Effect of Mach number on the non-equilibrium degree and
the degree of correlation.}
\end{figure}

\section{Conclusions}

In the paper the two-dimensional Richtmyer-Meshkov instability
system and the case combined with Rayleigh-Taylor instability are
simulated with a MRT discrete Boltzmann model, and the
nonequilibrium characteristics are systematically investigated.
The degrees of correlation between macroscopic nonuniformities and
various globally averaged nonequilibrium strength values in RM
instability are analyzed. Heat conduction plays a major role in
the degree of correlation. The greater the heat conduction, the
higher the degree of correlation. In the case combined with RT
instability, the collaboration and competition mechanisms of the
two instabilities are investigated, and the effects of gravity
field $g$ and Mach number on nonequilibrium are probed. (i) The
interface inversion process in RM instability system is affected
observably by the gravity field, and the regions for interface
inversion or not are obtained. (ii) The thermodynamic
nonequilibrium characteristics of the RT and RM instability
coexisting system with different gravity fields get a more
systematic understanding. (iii) The effect of Mach number on the
non-equilibrium degree is studied. With the increase of Mach
number, the non equilibrium degree of the system is increased
exponentially, and the degree of correlation almost exponentially
decreases with the increase of Mach number.

\section*{Acknowledgments}

We acknowledge support from the National Natural Science
Foundation of China (under Grant Nos. 11402138, 11475028 and
11772064), Science Challenge Project (JCKY2016212A501), the
Natural Science Foundation of Shandong Province (ZR2014EEP021) and
the Foundation for Outstanding Young Scientist in Shandong
Province (2012BSB01516).



\begin{thebibliography}{99}


\bibitem{rm1} R. D. Richtmyer, Taylor instability in
shock acceleration of compressible fluids, Commun. Pure Appl.
Maths. 13(2), 297 (1960)

\bibitem{rm2} E. E. Meshkov, Instability of the interface between
of two gases accelerated by a shock wave, Sov. Fluid Dyn. 4(5),
101 (1969)

\bibitem{rt1} L. Rayleigh, Investigation of the character of the
equilibrium of an incompressible heavy fluid of variable density,
Proc. London Math. Soc. s1-14(1), 170 (1882)

\bibitem{rt2} G. Taylor, The Instability of Liquid Surfaces when
Accelerated in a Direction Perpendicular to their Planes. I, P.
Roy. Soc. A 201(1065), 192 (1950)

\bibitem{PhysicsReports2017a} Y. Zhou, Rayleigh-Taylor and Richtmyer-Meshkov instability induced
flow, turbulence, and mixing. I, Physics Reports 720-722, 1-136 (2017).

\bibitem{PhysicsReports2017b} Y. Zhou, Rayleigh-Taylor and Richtmyer-Meshkov instability induced
flow, turbulence, and mixing. II, Physics Reports 723-725, 1-160 (2017).

\bibitem{Xu2016SC-review} A. Xu, G. Zhang, Y. Ying, C. Wang, Complex fields in heterogeneous materials
under shock: Modeling, simulation and analysis, Sci. China: Phys. Mech. \& Astron.
59(5), 650501 (2016)

\bibitem{Betti2006} R. Betti, and J. Sanz, Bubble acceleration in the ablative
Rayleigh-Taylor instability, Phys. Rev. Lett. 97(20), 205002
(2006)

\bibitem{Gupta2010} M. R. Gupta, L. Mandal, S. Roy, and M. Khan, Effect of magnetic field
on temporal development of Rayleigh-Taylor instability induced
interfacial nonlinear structure, Phys. Plasmas 17(1), 012306
(2010)

\bibitem{Sharma2010} P. K. Sharma, R. P. Prajapati, and R. K. Chhajlani, Effect of Surface
Tension and Rotation on Rayleigh-Taylor Instability of Two
Superposed Fluids with Suspended Particles, Acta Phys. Pol. A
118(4), 576 (2010)

\bibitem{Banerjee2011} R. Banerjee, L. K. Mandal, S. Roy, M. Khan, and M. R. Gupta, Combined
effect of viscosity and vorticity on single mode Rayleigh-Taylor
instability bubble growth, Phys. Plasmas 18(2), 022109 (2011)


\bibitem{Orlicz} G. C. Orlicz, B. J. Balakumar, C. D. Tomkins, K. P.
Prestridge, A Mach number study of the Richtmyer¨CMeshkov
instability in a varicose, heavy-gas curtain, Phys. Fluids 21(6),
064102 (2009)

\bibitem{Mostert} W. Mostert, V. Wheatley, R. Samtaney, and D. I. Pullin, Effects of
magnetic fields on magnetohydrodynamic cylindrical and spherical
Richtmyer-Meshkov instability, Phys. Fluids 27(10), 104102 (2015)


\bibitem{KM2005} K. O. Mikaelian, Rayleigh-Taylor and Richtmyer-Meshkov
instabilities and mixing in stratified cylindrical shells, Phys.
Fluids 17(9), 094105 (2005)

\bibitem{ML2007} M. Latini, O. Schilling, and W. S. Don, High-resolution simulations
and modeling of reshocked single-mode Richtmyer-Meshkov
instability: Comparison to experimental data and to amplitude
growth model predictions, Phys. Fluids 19(2), 024104 (2007)

\bibitem{tbl1} B. L. Tian, D. X. Fu and Y. W. Ma, Numerical investigation of Richtmyer-
Meshkov instability driven by cylindrical shocks, Acta Mech.
Sinica 22(1), 9 (2006)

\bibitem{tbl2} F. J. Gao, Y. S. Zhang, Z. W. He, B. L. Tian, Formula for growth rate of mixing
width applied to Richtmyer-Meshkov instability, Phys. Fluids
28(11), 114101 (2016)

\bibitem{chen2011} F. Chen, A. Xu , G. Zhang , Y. Li , Multiple-Relaxation-Time
Lattice Boltzmann Approach to Richtmyer-Meshkov Instability,
Commun. Theor. Phys. 55(2) 325 (2011)

\bibitem{LB2011} L. Biferale, F. Mantovani, M. Sbragaglia, A. Scagliarini, F. Toschi, R. Tripiccione,
Reactive Rayleigh-Taylor systems: Front propagation and
non-stationarity, Europhys. Lett. 94(5), 54004 (2011)

\bibitem{guo2013} G. J. Liu, Z. L. Guo, Effects of Prandtl number on mixing process in miscible
Rayleigh-Taylor instability: A lattice Boltzmann study, Int. J.
Numer. Method. H. 23(1), 176 (2013)

\bibitem{wang2016} L. F. Wang, W. H. Ye, J. F. Wu, Jie Liu, W. Y. Zhang, X. T. He,
A scheme for reducing deceleration-phase Rayleigh-Taylor growth in
inertial confinement fusion implosions, Phys. Plasmas 23(5),
052713 (2016)

\bibitem{IM2016} I. Maimouni, J. Goyon, E. Lac, T. Pringuey, J. Boujlel, X. Chateau,
P. Coussot, Rayleigh-Taylor Instability in Elastoplastic Solids: A
Local Catastrophic Process, Phys. Rev. Lett. 116(15), 154502
(2016)

\bibitem{Aschenbach1995} B. Aschenbach, R. Egger, J. Tromper, Discovery of
explosion fragments outside the Vela supernova remnant shock-wave
boundary, Nature 373, 587 (1995)

\bibitem{Balick2002} B. Balick and A. Frank, Shapes and Shaping of Planetary Nebulae,
Annual Review of Astronomy and Astrophysics, 40, 439 (2002)

\bibitem{Bradley2014} P. A. Bradley, The effect of mix on capsule yields as a function
of shell thickness and gas fill, Phys. Plasma 21(6), 062703 (2014)

\bibitem{Attal2015} N. Attal and P. Ramaprabhu, Numerical investigation of a single-mode
chemically reacting Richtmyer-Meshkov instability, Shock Waves
25(4), 307 (2015)

\bibitem{Matsumoto2013} J. Matsumoto and Y. Masada, Two-dimensional numerical study for
Rayleigh-Taylor and Richtmyer-Meshkov instabilitites in
relativistic jets, Astrophys. J. Lett. 772, 1 (2013)

\bibitem{Meshkov2013} E. E Meshkov, Some peculiar features of hydrodynamic instability development,
Phil Trans. R. Soc. A 371(2003), 20120288 (2013)

\bibitem{he2016} X. T. He, J. W. Li, Z. F. Fan, L. F. Wang, J. Liu, K. Lan, J. F.
Wu, A hybrid-drive nonisobaric-ignition scheme for inertial
confinement fusion, Phys. Plasmas 23(8), 082706 (2016)

\bibitem{xu2015aps} A. Xu, G. Zhang, Y. Ying, Progess of discrete Boltzmann modeling and simulation
of combustion system, Acta Phys. Sin. 64(18), 184701 (2015)

\bibitem{xu2016} A. Xu, G. Zhang, Y. Gan, Progress in studies on discrete Boltzmann modeling
of phase separation process, Mech. Eng. 38(4), 361 (2016)

\bibitem{Kinetic2018} A. Xu, G. Zhang and Y. Zhang, Discrete Boltzmann Modeling of Compressible Flows,
Chapter 2 in Kinetic Theory edited by G. Z. Kyzas and A. C. Mitropoulos, Rijeka: InTech, 2018.




\bibitem{ss2001} S. Succi, The Lattice Boltzmann Equation for Fluid Dynamics and Beyond, Oxford:
Oxford University Press, 2001

\bibitem{ss1992} R. Benzi, S. Succi, and M. Vergassola, The lattice Boltzmann equation:
Theory and applications, Phys. Rep. 222(3), 145 (1992)

\bibitem{xu2012} A. Xu, G. Zhang, Y. Gan, F. Chen, X. Yu, Lattice Boltzmann modeling and simulation
of compressible flows, Front. Phys. 7(5), 582 (2012)

\bibitem{Yeoman1995PRL} W. R. Osborn, E. Orlandini, M. R. Swift, J. M. Yeomans, J. R. Banavar,
Lattice boltzmann study of hydrodynamic spinodal decomposition, Phys. Rev. Lett. 75(22), 4031 (1995).

\bibitem{Fang-Qian2004PRE} H. Li, X. Lu, H. Fang, Y. Qian, Force evaluations in lattice boltzmann
simulations with moving boundaries in two dimensions, Phys. Rev. E 70(2), 026701 (2004).

\bibitem{Qin2005PRE} Y. Zhang, R. Qin, and D. Emerson, Lattice Boltzmann
simulation of rarefied gas flows in microchannels, Phys.
Rev. E 71(4), 047702 (2005)

\bibitem{Shan2016JFM} X. Shan, X. Yuan and H. Chen, Kinetic theory representation of hydrodynamics: a way beyond the Navier-Stokes equation, J Fluid Mech. 550, 413 (2006)

\bibitem{Shu2015JCP} Y. Wang, C. Shu, H. Huang, et al. Multiphase lattice Boltzmann flux solver for incompressible multiphase flows with large density ratio. J Comput. Phys. 280, 404 (2015)

\bibitem{Shu2007PRE} K. Qu, C. Shu and Y. Chew, Alternative method to construct equilibrium distribution functions in lattice-Boltzmann method simulation of inviscid
compressible flows at high Mach number. Phys. Rev. E 75, 036706
(2007)

\bibitem{Zhong2012PRE} C. Zhuo, C. Zhong and J. Gao, Filter-matrix lattice Boltzmann model for incompressible thermal flows, Phys. Rev. E 85,046703 (2012)

\bibitem{Zhong2015} K. Li, C. Zhong, A lattice Boltzmann model for simulation of compressible flows, Int. J. Numer. Meth. Fluids 77(6),334 (2015)

\bibitem{Zhang-Qin2005JSP} Y. Zhang, R. Qin, Y. Sun, R. W. Barber, and D. R.
Emerson, Gas flow in microchannels¨Ca lattice Boltzmann method approach, J. Stat. Phys. 121(1¨C2), 257
(2005).



\bibitem{yan2013} B. Yan, A. Xu, G. Zhang, Y. Ying, H. Li, Lattice Boltzmann model for combustion and detonation, Front. Phys. 8(1), 94 (2013)

\bibitem{gan2015} Y. Gan, A. Xu, G. Zhang, S. Succi, Discrete Boltzmann modeling of multiphase flows:
hydrodynamic and thermodynamic non-equilibrium effects, Soft
Matter 11(26), 5336 (2015)

\bibitem{xu2015pre} A. Xu, C. Lin, G. Zhang, Y. Li, Multiple-relaxation-time lattice Boltzmann
kinetic model for combustion, Phys. Rev. E 91(4), 043306 (2015)

\bibitem{Lin2016CNF} C. Lin, A. Xu, G. Zhang, Y. Li, Double-distribution-function discrete Boltzmann model for combustion, Combust. Flame 164,137(2016)

\bibitem{zhang2016CNF} Y. Zhang, A. Xu, G. Zhang, C. Zhu, and C. Lin, Kinetic modeling of detonation
and effects of negative temperature coefficient, Combust. Flame
173, 483 (2016)

\bibitem{Lin2017PRE} C. Lin, A. Xu, G. Zhang, K. Luo, Y. Li, Discrete Boltzmann modeling of Rayleigh-Taylor instability in two-component compressible flows, Phys. Rev. E 96, 053305 (2017)

\bibitem{lin2014} C. Lin, A. Xu, G. Zhang, Y. Li, S. Succi, Polar-coordinate lattice Boltzmann modeling
of compressible flows, Phys. Rev. E 89(1), 013307 (2014)

\bibitem{chen2014} F. Chen, A. Xu, G. Zhang, Y. Wang, Two-dimensional MRT LB model for compressible
and incompressible flows, Front. Phys. 9(2), 246 (2014)

\bibitem{lai2016} H. Lai, A. Xu, G. Zhang, Y. Gan, Y. Ying, S. Succi, Non-equilibrium thermohydrodynamic
effects on the Rayleigh-Taylor instability incompressible flow,
Phys. Rev. E 94(2), 023106 (2016)

\bibitem{chen2016} F. Chen, A. Xu, and G. Zhang, Viscosity, heat conductivity, and Prandtl number
effects in the Rayleigh-Taylor Instability, Front. Phys. 11(6),
114703 (2016)

\bibitem{Lin-2018-CAF} C. Lin, K. Luo, MRT discrete Boltzmann method for compressible exothermic reactive flows, Computers and Fluids 166, 176 (2018).

\bibitem{Lin-2017-SR} C. Lin, K. Luo, L. Fei, S. Succi, A multi-component discrete Boltzmann model for
nonequilibrium reactive flows, Sci. Rep. 7, 14580 (2017).

\bibitem{Xu-2018-FoP} A. Xu, G. Zhang, Y. Zhang, P. Wang, Y. Ying, Discrete Boltzmann model for implosion and explosion related compressible flow with spherical symmetry, Front. Phys.(2018,in press);  arXiv:1803.03117.




\bibitem{ZYD-FoP2018} Y. Zhang, A. Xu, G. Zhang, Z. Chen, P. Wang, Discrete ellipsoidal statistical BGK model and Burnett equations,  Front. Phys. 13(3), 135101 (2018).

\bibitem{Gan2018} Y. Gan, A. Xu, G. Zhang, Y. Zhang, S. Succi,
Discrete Boltzmann trans-scale modeling of high-speed compressible flows,  arXiv:1801.04522

\bibitem{ZYD-2018} Y. Zhang, A. Xu, G. Zhang, Z. Chen, P. Wang, Discrete Boltzmann method for
nonequilibrium flows: based on Shakhov model, (Submitted to J. Fluid Mech.)




\bibitem{kw2016} H. Liu, W. Kang, Q. Zhang, Y. Zhang, H. Duan, and X. T. He, Molecular dynamics
simulations of microscopic structure of ultra strong shock waves
in dense helium, Front. Phys. 11(6), 115206 (2016)

\bibitem{kw2017} H. Liu, Y. Zhang, W. Kang, P. Zhang, H.L. Duan and X.T. He, Molecular dynamics
simulation of strong shock waves propagating in dense deuterium,
taking into consideration effects of excited electrons. Phys. Rev.
E 95(7), 023201 (2017)

\bibitem{kw2017b} H. Liu, W. Kang, H. Duan, P. Zhang, X. He,
  Recent progresses on numerical investigations of microscopic structure of strong shock waves in fluid,
  Scientia Sinica Physica Mechanica $\&$ Astronomica 47(7), 070003, 2017.


\bibitem{Meng2013JFM}
J. Meng, Y. Zhang, N. G. Hadjiconstantinou, G. A. Radtke, and X. Shan, Lattice ellipsoidal statistical BGK
model for thermal non-equilibrium flows, J. Fluid Mech. 718, 347 (2013)

\end{thebibliography}
\end{document}